\documentclass[12pt,draftclsnofoot,onecolumn]{IEEEtran}

\usepackage{latexsym}
\usepackage{graphicx}
\usepackage{epsfig}
\usepackage{subfigure}
\usepackage{array}
\usepackage{amsmath}
\usepackage{amssymb}
\usepackage{color, soul}
\usepackage{amsthm}
\usepackage{enumerate}
\usepackage{algpseudocode}
\usepackage{algorithm}

\newcommand{\bpp}{\textbf p}

\newcommand{\bR}{\textbf R}
\newcommand{\bA}{\textbf A}
\newcommand{\bB}{\textbf B}
\newcommand{\bD}{\textbf D}

\newcommand{\bU}{\textbf U}

\newcommand{\bW}{\textbf W}

\newcommand{\bH}{\textbf H}

\newcommand{\bh}{\textbf h}
\newcommand{\bx}{\textbf x}

\newcommand{\bI}{\textbf I}

\newcommand{\by}{\textbf y}

\newcommand{\bxi} {\boldsymbol{\xi}}

\DeclareMathOperator{\e}{\mathbb{E}}
\DeclareMathOperator{\tr}{tr}
\newcommand{\bp} {\begin{proof}}
\newcommand{\ep} {\end{proof}}

\makeatletter
\newcommand*{\rom}[1]{\expandafter\@slowromancap\romannumeral #1@}
\makeatother

\newtheorem{thm}{Theorem}

\newtheorem{lemma}{Lemma}
\newtheorem{prop}{Proposition}
\theoremstyle{definition}

\theoremstyle{definition}
\newtheorem{definition}{Definition}

\def\bal#1\eal{\begin{align}#1\end{align}}
\begin{document}
 \bstctlcite{IEEEexample:BSTcontrol}
\date{}

\title{Optimal Power Assignment for MIMO Channels Under Joint Total and Per-Group Power Constraints}

\author{M. Khojastehnia\thanks{Mahdi Khojastehnia, Ioannis Lambadaris and Ramy H. Gohary are with the Department of Systems and Computer Engineering, Carleton University, Ottawa, ON, Canada, e-mail: khojastehnia.m@gmail.com, \{ioannis,gohary\}@sce.carleton.ca},
\and I. Lambadaris,
\and R.H. Gohary,
\and S. Loyka\thanks{Sergey Loyka is with the School of Electrical Engineering and Computer Science, University of Ottawa, Ottawa, ON, Canada, e-mail: Sergey.Loyka@uottawa.ca}
}

\maketitle

\vspace*{-2\baselineskip}

\begin{abstract}
    In this paper we consider a 
    communication system with one transmitter and one receiver. The transmit antennas are partitioned into disjoint groups, and each group must satisfy an average  
    power constraint  in addition to the standard overall one. 
    The optimal power allocation (OPA) for the transmit antennas is obtained for the following cases: (i) fixed multiple-input multiple-output (MIMO) orthogonal channel, (ii) i.i.d. fading MIMO orthogonal channel, and (iii) i.i.d. Rayleigh fading multiple-input single-output (MISO) and MIMO channels. The channel orthogonality is encountered in the practical case of the massive MIMO channel under favorable propagation conditions. The closed-form solution to the OPA for a fixed channel is found using the Karush-Kuhn-Tucker (KKT) conditions and it is similar to the standard water-filling procedure while the effect of the per-group average power constraint is added. For a fading channel, an algorithm is proposed to give the OPA, and the algorithm's convergence is proved via a majorization inequality and a Schur-concavity property.

\end{abstract}


\section{Introduction}
Multiple-input multiple-output (MIMO) or multi-antenna systems give notable benefits for wireless communications including higher data rate and interference reduction with the same usage of bandwidth and power \cite{Biglieri-07}. In industry applications, MIMO systems exist in wireless communication standards such as the long-term evolution (LTE) \cite{Cox-14} and is considered as a key technology for 5G in the form of massive MIMO systems \cite{Marzetta-15}\cite{Shafi-17}. 

The capacity and optimal signaling for a fixed and Rayleigh fading channels under the total power constraint (TPC) are known. For a fixed channel, the optimal beam direction are on the channel eigenmodes and the optimal power allocation is via the water-filling procedure \cite{Telatar-99}. For the latter case and when the receiver (Rx) knows the channel and the transmitter (Tx) knows the channel distribution, the independent uniform power allocation is optimal \cite{Tse-05}. The TPC is motivated by limited power budget at the power supply. TPC affects battery life  \cite{Wang-19}\cite{Loyka-17} and also it should be limited because of the environmental safety \cite{Khoshnevisan-12}\cite{khojastehnia-thesis-19}. Per-group power constraint (PGPC) is motivated since each group of antennas are connected to their amplifier in which it has a limited power. The consideration of the joint TPC and PGPC constraints make the problem of finding the capacity more complicated as optimal Tx covaraince and the channel Gram matrix cannot have orthogonal eigenvectors in general. The capacity and optimal Tx covaraince under the joint power constraints are not known in general, except for special cases, e.g. MISO channels and full-rank solution under the joint TPC and per-antenna power constraint\footnote{Per-antenna power constraint (PAPC) is a special case of PGPC for which each group has a single antenna.} for fixed channel; and MISO channel under the PAPC alone for a Rayleigh fading channel \cite{Vu-11}.

\subsection{Related research}
\textit{Research on fixed channel}: For a fixed MISO channel under per-antenna power constraint (PAPC) alone, it is shown that beamforming (i.e. rank-1 solution) for the Tx covariance is optimal \cite{Vu-11}. This is also the case for MISO channel under combined TPC and PAPC \cite{Loyka-17}, under combined TPC, PAPC, and PGPC \cite{Chaluvadi-19}, and MISO wiretap channel under the joint TPC and PAPC with 2 Tx antennas \cite{Cao-17}. The difference between these cases is in the beamforming vector entries. As an example under TPC, they are dependent on the channel gain matrix. Furthermore, under PAPC, they are dependent on the corresponding constraint parameters. For the case of two users, \cite{Boccardi-06} studies the sum capacity using the water-filling procedure and gives the closed-form solution for it. Under the joint TPC and PAPC and for two antenna transmitters, the closed-form solution for the optimal signaling is obtained in \cite{Cao-16}. References \cite{Vu-11}, \cite{Loyka-17} give the optimal Tx covariance for a fixed MISO channel under the PAPC alone and for the joint TPC and PAPC respectively. The full rank solution for a fixed MIMO channel under the joint TPC and PAPC constraints is obtained in \cite{Loyka-19}, for which the optimal power allocation (OPA) for Tx antennas are the minimum of PAPC and OPA under the TPC alone. Although the optimal Tx covariance under PGPC and combined TPC and PGPC is still unknown in general, \cite{Vu-11-2} obtains the optimal signaling for a MIMO channel under PAPC as a function of dual variables and an iterative algorithm is proposed in order to determine optimal values. For the case of multi user setting and under PAPC, \cite{Park-10} proposes a numerical algorithm to compute the optimal signaling for the weighted sum rate maximization problem. As well, under PAPC and for multiple access MIMO channel, \cite{Zhu-12} proposes a numerical algorithm in order to compute the sum capacity. Under TPC and PAPC, \cite{Cao-17-2} gives an algorithm to compute the optimal Tx covariance for a MIMO channel which covers the more general case compared to the algorithm proposed in \cite{Cao-16} for MISO channels. The case of combined TPC, PAPC, and PGPC is considered in \cite{Chaluvadi-19}. This paper examines the case of full-rank channel and full rank Tx covariance. The off-diagonal entries of optimal Tx covariacne are given in closed-forms and an algorithm is proposed to compute its diagonal entries. For the case of 2 Tx antennas, the closed form expression for the optimal Tx covariane is obtained and an algorithm is proposed to compute the optimal signaling for the general MIMO channel based on projected gradient descent approach. The correctness of the proposed algorithm is also verified by the CVX tool \cite{CVX-18}.

\textit{Research on fading channel}: In a MIMO channel with i.i.d. Rayleigh fading channel, where the Rx knows the channel and Tx knows the statistical channel realization, the optimal Tx covariance is shown to be diagonal \cite{Telatar-99}. This is also the case in a Rayleigh MISO channel under PAPC alone. However, for the former one, uniform power allocation is optimal for Tx antennas and for the latter one, the OPA for Tx antennas is equal to PAPC \cite{Vu-11}. The case of a Gaussian fading MISO channel under the PAPC is considered in \cite{Maamari-14} and the ergodic capacity is achieved where full channel state information (CSI) is known at both Tx and Rx. In this case, beamforming signaling is optimal. In \cite{Loyka-17}, the case of right unitary-invariant fading channel\footnote{In the right unitary-invariant fading channel model, Rx correlations can be existed while it is not the case for Tx.} is considered under the combined TPC and PAPC constraints. It is shown that isotropic signaling is optimal. In \cite{Favano-20}, the channel capacity under PAPC is analyzed for the case when both Tx and Rx have knowledge of the channel distribution. For the channel capacity, the upper-bound and lower-bound are determined. It is further shown that the capacity gap tends to zero by letting the PAPC approach infinity. The Rician/Lognormal fading distributed MIMO channel is considered in \cite{Wang-20} where Tx only knows the channel distribution. The lower-bound of the ergodic capacity is obtained and numerical results show that this lower-bound is tight for most SNR values. In \cite{Qiu-20}, the downlink total power minimization problem is studied under PAPC and downlink rate constraints. The considered model is spatially correlated Rayleigh fading cell-free massive MIMO channel and a numerical algorithm is proposed to compute the optimal solution for the corresponding optimization problem.   
\subsection{Contribution of our work}
The optimal transmit covariance of the fixed and fading channel under the joint total power and per-group power constraints is not known in general, except for some special cases \cite{Loyka-17}\cite{Vu-11}\cite{Cao-17} \cite{Cao-17-2}\cite{Cao-16}\cite{Loyka-19}\cite{Vu-11-2}\cite{Zhu-12}. Although there are some numerical algorithms in the literature in order to compute the optimal transmit covariance \cite{Chaluvadi-19}\cite{Qiu-20}\cite{Vu-11-2}, those algorithms are in general computationally demanding, since their complexity grows with $O(n^6)$ where $n$ is the number of transmit antennas. The contribution of our work is as follows:
\begin{itemize}
\item The closed-form solution for optimal transmit covariance under the joint TPC and PGPC is obtained for an orthogonal fixed MIMO channel.
\item A finite-step numerical algorithm derived from the mathematical analysis for the optimal transmit covariance matrix under the joint TPC and PGPC for an i.i.d. Rayleigh fading MIMO channel and an orthogonal i.i.d. fading MIMO channel. As well, for the particular cases of i.i.d. Rayleigh fading MISO channel and i.i.d. Rayleigh fading MIMO channel with the two TX antennas.   
\end{itemize}
In our first contribution, we rely on the KKT conditions in order to solve the optimization problem, and for the second contribution, we rely on the Schur-concavity property of the objective function.

\emph{Notations}: Complex-valued matrices and vectors are denoted by bold capitals and bold lower-case letters, respectively; $|\bH|$ and $\bH^+$ denote the determinant and Hermitian conjugate of $\bH$; $\bR \ge 0$ means that $\bR$ is positive semi-definite; $\bx \succ \by$ means that the vector $\by$ is majorized by the vector $\bx$; $\bD(r_i)$ is a diagonal matrix with diagonal entries $r_i$; $(\bR)_{ij}$ is the $ij$-th element of $\bR$; $\bI$ is the identity matrix with appropriate dimension; $(x)_+ = \max\{x,0\}$.

\section{System Model and Capacity}
The output of a discrete-time fixed and memoryless Gaussian MIMO channel is modeled by:
\bal
\label{eq.MIMO-model}
\by = \bH\bx+\bxi,
\eal
where $\bx_{m \times 1}$ and $\by_{n \times 1}$ are the transmitted and received complex signal vectors; $\bxi_{n \times 1}$ is the i.i.d. AWGN; $\bH_{n\times m}$ is the channel gain matrix from the transmit antennas to the receive antennas; $n$, $m$ are the number of antennas at the receiver and transmitter respectively.

\subsection{Fixed MIMO channel}
When the MIMO channel $\bH$ is fixed and completely known to both the transmitter and the receiver, the channel capacity $C^*$ [nat/s/Hz] is given by \cite{Biglieri-07}\cite{Tse-05}:
\bal
\label{eq.fixed-capacity}
C^* = \max_{\bR \in S_R}\ln|\bI+\bW\bR|
\eal
where $\bR = \e\{\bx\bx^+\}$ is the transmit (Tx) covariance matrix, $\bW  = \bH^+\bH$ is the channel Gram matrix, and $S_R$ is the feasible set containing the constraints on the Tx covariance matrices. 

For the case of the total power constraint only, the constraint set is $S_R = \{\bR: \bR \ge 0, \tr\bR \le P_T\}$ where $P_T >0$ is the maximum total Tx power. The optimal Tx covariance matrix is on the eigenvectors of the channel Gram matrix $\bW$, i.e. the eigenvectors of the optimal Tx covariance matrix is orthogonal to the eigenvectors of the channel Gram matrix $\bW$. In addition, the optimal power allocations for those beam vectors is obtained by the water-filling solution \cite{Telatar-99}, which is:
\bal
\label{eq.WF}
p^*_{i,WF} = \big(\mu_{WF}^{-1}-\lambda^{-1}_i(\bW)\big)_+, \quad i = 1,\ldots,m,
\eal 
where $p^*_{i,WF}$ is the optimal power allocation (OPA) for the $i$-th beam vector, $\lambda_i(\bW)$ is the $i$-th eigenvalue of the channel Gram matrix $\bW$. Both $p^*_{i,WF}$ and $\lambda_i(\bW)$ are in decreasing order, $\mu_{WF}$ is the water-level and can be found by the total power constraint $P_T$.

Under the joint total power constraint (TPC) and the per-group power constrain (PGPC), the feasible set becomes:
\bal
\label{eq.feasible-set}
S_R = \bigg\{\bR: \bR \ge 0,\ \tr\bR \le P_T,\ \sum_{i \in I(j)}(\bR)_{ii} \le P_j,\ j=1,\ldots,s \bigg\},
\eal
where $(\bR)_{ii}$ is the $i$-th diagonal entry of $\bR$ and gives the corresponding allocated power to the $i$-th Tx antenna; $I(j)$ is the set of antennas in the $j$-th group; $P_j>0$ is the $j$-th per-group power constraint (PGPC); the sets $I(i)$ and $I(j)$ are disjoint for all $i \ne j$ and $I(1)\cup...\cup I(s) = \{1,...,m\}$; $s$ is the number of antenna groups. The optimal Tx covariance matrix is not known in general for this case, except for some special cases (e.g. MIMO orthogonal and MISO channels under the joint total and per-antenna power constraints \cite{Khojastehnia-19}\cite{Loyka-17}). In this paper, for the case of fixed orthogonal channel, we present the solution to the problem in \eqref{eq.fixed-capacity} under the joint TPC and PGPC constraint set \eqref{eq.feasible-set}. Our approach has two differences compared to the previous studies \cite{Loyka-17}\cite{Vu-11}  \cite{Cao-17}\cite{Cao-17-2}\cite{Cao-16}\cite{Loyka-19}\cite{Vu-11-2}\cite{Zhu-12}: a) the optimization problem is considered under the joint TPC and per-group power constraints, contrary to the past studies which focused on TPC and per-antenna power constraints only, b) we derived the closed-form solution for $\bR^*$ under the joint TPC and PGPC and thus, avoiding the need to resort to high demanding computational algorithms for $\bR^*$. 

\subsection{Fading MIMO channel}
In this part of our work, we assume that the channel is completely known to the receiver and the transmitter only knows the statistical distribution of the channel. The channel capacity $C^*$ is given by \cite{Biglieri-07}\cite{Tse-05}:
\bal
\label{eq.fading-capacity}
C^* = \max_{\bR \in S_R}\e_{\bH}\big\{\ln|\bI+\bH\bR\bH^+|\big\}
\eal
Under the TPC only and for the case of the Rayleigh distribution, it is shown that isotropic signaling maximizes the ergodic rate, that is the optimal Tx covariance matrix $\bR^*$ is diagonal and the power is equally allocated to all Tx antennas \cite{Telatar-99}, i.e. $\bR^* = \frac{P_T}{m}\bI$.

Under the joint total and the per-group power constraints, the optimal solution for the Tx covariance matrix is not known in general, except for few special cases: (i) MISO channel under the per-antenna power constraint (PAPC) \cite{Vu-11}, (ii) MISO channel under the joint TPC and the PAPC, where PAPC are the same for all Tx antennas \cite{Loyka-17}. It has been shown for both cases the optimal Tx covariance is diagonal. For the former one, the optimal power allocation for transmit antennas (i.e. the diagonal entries of the optimal Tx covariance matrix) is equal to the PAPC values, and for the latter one is the minimum of PAPC values and $\frac{P_T}{m}$. Hence, independent signaling is optimal for both cases. In our work, we consider an orthogonal fading channel and present the algorithm to obtain $\bR^*$. The orthogonality of the MIMO channels is observed in the study of the practical massive MIMO channel under the \textit{favorable} propagation channel \cite{Marzetta-16}. Both theoretical-based and measurement-based results verify channel orthogonality in massive MIMO channels \cite{Masouros-15}\cite{Gao-15}\cite{Gauger-15}\cite{Hoydis-12}\cite{Gao-15-2}\cite{Harris-17-2}\cite{Yang-17}\cite{Martinez-18}. Also, such orthogonal models, can be used for OFDM communication in which the power constraint are considered for the orthogonal narrowband frequency tones \cite{Gohary-09}. Then, we extend this approach to obtain $\bR^*$ for the Rayleigh fading MIMO channel. This approach is based on the theoretical solution for the optimal Tx covariance matrix $\bR^*$ of the problem in \eqref{eq.fading-capacity}.
     
\section{The Capacity Derivation of the Constant MIMO Orthogonal Channel}
\label{sec.fixedChannel}
In this section, we consider the orthogonal MIMO channel where the columns vectors of $\bH$ are orthogonal to each other, i.e. $\bH\bH^+=\bW = \bD(g_i)$ and $g_i$ is the $i$-th diagonal entry of $\bW$. The next theorem gives the closed-form solution to \eqref{eq.fixed-capacity} under the joint TPC and PGPC constraints. To this end, let $p_i$ be the $i$-th diagonal entry of $\bR$.

\begin{thm}
\label{thm.R*-fixed}
Consider the feasible set in \eqref{eq.feasible-set} and let the $i$-th transmit antenna belong to the $j$-th group. The optimal transmit covariance matrix $\bR^*$ in \eqref{eq.fixed-capacity} under the joint TPC and PGPC constraints for the orthogonal MIMO channel is diagonal, i.e. $\bR^*=\bD(p^*_i)$, where $p^*_i$ is the optimal power allocation for the $i$-th transmit antenna belonging to the $j$-th group given by:
\bal
\label{eq.p*_i.fixed}
p^*_i = \big((\mu+\lambda_j)^{-1}-g_i^{-1}\big)_+
\eal 
where $\mu \ge 0$ and $\lambda_j \ge 0$ are the Lagrange dual variables responsible for the total power constraint and the $j$-th group power constraint respectively (following Proposition \ref{prop.mu.lambda_j} for details pertaining to $\mu, \lambda_j$).

Since both $\bR^*$ and $\bW$ are diagonal, the channel capacity can be obtained as follows:
\bal
\label{eq.C^*.fixed}
C^* = \sum_{i=1}^m\ln(1+g_ip^*_i).
\eal

\begin{proof}
The proof can be found in Appendix A.
\end{proof}

\end{thm}

From the theorem we conclude that independent signaling is optimal for the orthogonal channel under the joint TPC and PGPC constraints. Furthermore, the effect of the PGPC on the optimal power allocation $p^*_i$ is determined by $\lambda_j$. When the PGPC constraint is removed, i.e. considering the TPC alone, the solution to $p_i$ coincides with the water-filling solution in \eqref{eq.WF}. Similar to the water-filling solution, stronger streams (i.e. corresponding to larger $g_i$) get more power compared to the weaker ones, that is $g_i \ge g_j$ implies $p^*_i \ge p^*_j$.

The next Proposition shows how dual variables $\mu, \lambda_j$ can be derived.

\begin{prop}
\label{prop.mu.lambda_j}
If $\sum_{j=1}^s P_j \le P_T$, then TPC is inactive and $\mu = 0$. Otherwise, the dual variable $\mu$ satisfies the following non-linear equation: 
\bal
\label{eq.mu.find}
\sum_{j=1}^s \min \bigg\{P_j, \sum_{i \in I(j)}(\mu^{-1}-g^{-1}_i)_+\bigg\} = P_T.
\eal
Once $\mu$ is determined, the dual variables $\lambda_j$ can be derived from the following non-linear equation:
 \bal
 \label{eq.lambda_j.find}
\sum_{i \in I(j)}\big((\mu+\lambda_j)^{-1}-g_i^{-1}\big)_+ = \min \bigg\{P_j, \sum_{i \in I(j)}(\mu^{-1}-g^{-1}_i)_+\bigg\},\quad j=1,\ldots,s,
\eal

\begin{proof}
The proof can be found in Appendix B.
\end{proof}

\end{prop}  

From this proposition, we observe that the total optimal power allocation (OPA) for the $j$-th group is $\min \bigg\{P_j, \sum_{i \in I(j)}(\mu^{-1}-g^{-1}_i)_+\bigg\}$. This expression consists of two terms: The first one is the $j$-th PGPC while the second one is the OPA found from the water-filling procedure. However, the dual variable responsible for the TPC under the joint TPC and PGPC constraints may not be the same with that under the the TPC alone. 

To obtain the dual variables $\mu, \{\lambda_{j}\}$ from the above non-linear monotonic equations \eqref{eq.mu.find},\eqref{eq.lambda_j.find}, one can use the bisection algorithm. In each step of this iterative algorithm, the bound of the problem solution becomes tighter until it gets to the desired interval width for the problem solution \cite{Boyd-04}. To use this algorithm, the lower and upper bounds for dual variables are needed. They are given in the next proposition. Note that $\{\lambda_{j}\}$ are not coupled.

\begin{prop}
\label{prop.bounds-dual}
The dual variable $\mu$ satisfies:
\bal
\label{eq.bound.mu}
0 \le \mu \le \max_{i} g_i.  
\eal
Once $\mu$ is determined, if $\min \bigg\{P_j, \sum_{i \in I(j)}(\mu^{-1}-g^{-1}_i)_+\bigg\} = 0$, then $\lambda_j = \max_{i}g_i$. Otherwise,
\bal
\label{eq.bound.lambda_j}
0 \le \lambda_j \le \max_{i \in I(j)} g_i. 
\eal

\begin{proof}
For $\mu$, the lower bound is due to the dual feasible constraint $\mu \ge 0$. The upper bound follows from the fact that at least one stream is active in \eqref{eq.mu.find}, i.e. $(\mu^{-1}-g^{-1}_i)_+$ is positive for at least one $i$. 

For $\lambda_j$, if $\min \bigg\{P_j, \sum_{i \in I(j)}(\mu^{-1}-g^{-1}_i)_+\bigg\} = 0$, then \eqref{eq.lambda_j.find} implies that $\lambda_j$ can be any value such that $\lambda_j \ge g_i - \mu$, where $\max_{i}g_i$ is in this interval. Otherwise, if $\min \bigg\{P_j, \sum_{i \in I(j)}(\mu^{-1}-g^{-1}_i)_+\bigg\} > 0$, then $(\mu^{-1}-g^{-1}_i)_+$ is positive for at least one $i \in I(j)$, where it gives the upper bound of $\lambda_j$. The lower bound of $\lambda_j$ is due to the dual feasible constraint $\lambda_j \ge 0$
\end{proof}

\end{prop}

\section{The Capacity Derivation of the Fading MIMO Orthogonal Channel}
\label{sec.fadingChannel}
We will now consider the case of the fading MIMO orthogonal channel, and proceed with an argument similar to the one used in Theorem \ref{thm.R*-fixed}. It follows directly that the optimal Tx covariance $\bR^*$ is diagonal. Hence, the channel capacity in \eqref{eq.fading-capacity} can be expressed as follows:
\bal
\label{eq.C^*-fading-orthogonal}
C^*= \max_{p_i \in S_P}C(\{p_i\})
\eal
where
\bal
\label{eq.C(p_i)}
C(\{p_i\}_{i=1,2,\ldots,m}) = \e_{g_i}\bigg\{\sum_{i=1}^m\ln(1+g_ip_i)\bigg\},
\eal
where $\{g_i\}$ are positive i.i.d. random variables (there is no assumption on the distribution of $g_i$ in this section). The feasible set $S_P$ is obtained from $S_R$ in \eqref{eq.feasible-set} and it is as follows: 
\bal
\label{eq.S_P}
S_P =\Bigg\{\{p_i\}: p_i \ge 0, \sum_{i=1}^mp_i \le P_T, \sum_{i \in I(j)}p_i \le P_j, j=1,\ldots,s \Bigg\}.
\eal
The objective function $C(\{p_i\})$ is symmetric as it follows from the fact that $\{g_i\}$ are i.i.d. random variables. Furthermore, this function is concave. Under the sum constraint (i.e. TPC), this function is maximized when all arguments are the same \cite{Tse-05}. Since $C(\{p_i\})$ is (monotonically) increasing in $p_i$, the OPA for $p_i$ is $p^*_1=p^*_2=...=p^*_m=\frac{P_T}{m}$. Using this argument, for OPA under the joint TPC and PGPC constraints, we can easily show that antennas belonging to a certain group are assigned identical power. 

Under the joint TPC and PGPC constraints, the optimal Tx covariance $\bR^*$ is not known in general. Since the optimal Tx covariance is diagonal, we propose Algorithm 1 that obtains the OPA and hence $\bR^*$. It is worth mentioning that this algorithm gives the exact power assignment in a finite number of steps (see Proposition \ref {prop.algorithm}). The main idea of this approach is to identify all active group constraints (please check the details in the following paragraph). To do so, the average power with respect to the total power constraint is compared with the average power for each group power constraint. In an active group, the latter is smaller than the former. Furthermore, antennas in a certain active group will get the same amount of power. Once all active group constraints are found, the rest of the group power constraints are inactive and hence, the antennas in the remaining groups get identical powers.   

\begin{algorithm}[H]
\caption{OPA for the i.i.d. fading orthogonal channel under the joint TPC and PGPC}
\begin{algorithmic}
\Require All known variables in $S_P$, see \eqref{eq.MIMO-model}, \eqref{eq.feasible-set}.
\State \emph{\% Initialization}
\State $I = \{1,2,...,m\}$: all the Tx antenna indices, $I(j)$: set of the Tx antenna indices in the $j$-th group, $P=P_T, L=m$
\State \emph{\% Test whether there exists at least one active group by comparing $\frac{P}{L}$ (average power with respect to the TPC) and $\frac{P_j}{|I(j)|}$ (average power with respect to the $j$-th PGPC) for all $I(j) \in I$.}

\While {$\frac{P}{L} \ge \frac{P_j}{|I(j)|}$ for at least one $I(j)\in I$}

\For{each $I(j) \in I$} 

\If {$\frac{P}{L} \ge \frac{P_j}{|I(j)|}$} 

\State \emph{\% OPA for the transmit antennas belonging to the $j$-th active group}
\State $p^*_i = \frac{P_j}{|I(j)|}$ for $i \in I(j)$
\State \emph{\% Update remaining total power, number of antennas, and number of groups}
\State $P \leftarrow P-P_j$
\State  $L \leftarrow L-|I(j)|$
\State $I \leftarrow I\setminus I(j)$

\EndIf
\EndFor
\EndWhile

\For{each $I(j) \in I$} 
\State \emph{\% OPA for the transmit antennas in all inactive groups}
\State  $p^*_i = \frac{P}{L}$ for $i \in I$
\EndFor

\end{algorithmic}
\label{algorithm.OPA}
\end{algorithm}

Let $|I(j)|$ denote the number of antennas (elements) in the $j$-th group. The $j$-th group power constraint is active if $\sum_{i \in I(j)}p_i  = P_j$ (i.e. this group of antennas get the maximum allowed power), and it is inactive if $\sum_{i \in I(j)}p_i  < P_j$. Also, $P$ and $L$ (introduced in Algorithm 1 that follows) denote the remaining total power and remaining antennas in each step, respectively. In this algorithm, first the active groups are found by comparing  $\frac{P}{L}$ (average power with respect to $P_T$) with $\frac{P_j}{|I(j)|}$ (average power with respect to the $j$-th PGPC). If the former becomes greater than the latter, then the algorithm considers the $j$-th PGPC to be active. Then, certain number of active groups is determined and the power is uniformly allocated to antennas in each active group. For the remaining antennas, this comparison continues until additional active PGPC's are found. These iterations terminate when $\frac{P}{L}$ becomes smaller than $\frac{P_j}{|I(j)|}$. At this step, all remaining PGPC's are inactive and the residual power is uniformly allocated to the rest of the antennas. From this algorithm, we observe that  the antennas in active groups get less power compared to those in inactive groups.

In the next proposition, we show that this algorithm provides the optimal solution for the problem in \eqref{eq.C^*-fading-orthogonal}.
\begin{prop}
\label{prop.algorithm}
Consider the fading orthogonal MIMO channel under joint TPC and PGPC constraints. Algorithm \ref{algorithm.OPA} gives the optimal power allocation for $p_i$ for the case when the diagonal entries of the channel Gram matrix are i.i.d. random variables.

\begin{proof}
The proof can be found in Appendix C.
\end{proof}

\end{prop}  

We now present two special OPA cases for the i.i.d. fading orthogonal channel under the joint constraints: \\
\textit{Case 1}: The average power with respect to the total power constraint is less than the average power with respect to PGPC for all groups, i.e. $\frac{P_T}{m} \le \frac{P_j}{|I(j)|}$ for all $j$. Then:
\bal
\label{eq.special1}
C^* &\le C_{TPC} \notag \\
& =C\Big(\big\{p_i=\frac{P_T}{m}\big\}\Big),
\eal
where $ C_{TPC}$ is the capacity under the TPC alone, and the first inequality follows from the fact that the constraint set for $C^*$ is a subset of the constraint set of $C_{TPC}$. The upper-bound is achieved for $p^*_i = \frac{P_T}{m}$. Note that this OPA is in the feasible set since $\frac{P_T}{m} \le \frac{P_j}{|I(j)|}$ for all $j$.\\
\textit{Case 2}: The total power constraint is greater than the sum of PGPCs, i.e. $P_T \ge \sum_{j=1}^s P_j$. It follows that:
\bal
\label{eq.special1}
C^* &\le C_{PGPC}, 
\eal
where $ C_{PGPC}$ is the capacity under the PGPC alone. The first inequality is due to expanding the feseable set for $C_{PGPC}$. The upper-bound is achieved for $p^*_i = \frac{P_j}{|I(j)|}$ and $i \in |I(j)|$, which also is the OPA under the PGPC only.

\subsection{The Capacity of the Rayleigh fading MISO Channel}
In MISO channels, there is only one antenna at the receiver, and the channel gain matrix becomes a row vector, i.e. $\bH_{1\times m} = \bh$. In this case, the optimization problem for the capacity in \eqref{eq.fading-capacity} can be expressed as follows:
\bal
\label{eq.fading-capacity-MISO}
C^* = \max_{\bR \in S_R}\e_{\bh}\big\{\ln(1+\bh\bR\bh^+)\big\}
\eal
We assume that the entries of $\bh$ are zero mean i.i.d. complex circularly Gaussian random variables with unit variance (Rayleigh fading). Following the procedure in Appendix C of \cite{Vu-11}, we can show that $\bR^*$ is a diagonal matrix under the combined TPC and PGPC constraints. The proof relies on the fact that Rayleigh fading has a symmetric distribution. The optimization problem in \eqref{eq.fading-capacity-MISO} can be written as follows:
\bal
\label{eq.fading-capacity-MISO-diagonal}
C^* = \max_{\bR \in S_P}\e_{g_i}\bigg\{\ln(1+\sum_{i=1}^mg_ip_i)\bigg\}
\eal
where $g_i = |h_i|^2$. In the next proposition, we show that Algorithm \ref{algorithm.OPA} gives the OPA for this problem.

\begin{prop}
\label{prop.MISO-fading}
For the optimization problem in \eqref{eq.fading-capacity-MISO-diagonal} and the i.i.d. Rayleigh fading channel, algorithm \ref{algorithm.OPA} gives the optimal solution for $p_i$.

\begin{proof}
Using a similar argument as in Lemma \ref{lemma.Schur-concave}, we can show that the objective function in \eqref{eq.fading-capacity-MISO-diagonal} is Schur-concave since it is symmetric and concave. The rest of the proof follows the one given in Proposition \ref{prop.algorithm}.
\end{proof}

\end{prop}

\subsection{The Capacity of the Rayleigh Fading $2 \times n$ Channel}
Let us consider the i.i.d. Rayleigh fading MIMO channel with 2 transmit antennas and $n$ receive antennas. It implies that $\bR$ is 2 by 2 matrix. The next proposition gives the OPA for problem \eqref{eq.fading-capacity-MISO}.

\begin{prop}
\label{prop.2.n.MIMO}
For the i.i.d. Rayleigh fading channel and the optimization problem in \eqref{eq.fading-capacity} where $\bR$ is a 2 by 2 matrix, it follows that $\bR$ is diagonal, i.e. $\bR = \bD(p1,p2)$, and Algorithm \ref{algorithm.OPA} gives the optimal solution for $p_i$.

\begin{proof}
Please refer to Appendix D.
\end{proof}

\end{prop}  

\subsection{The Capacity of the Rayleigh Fading MIMO Channel}
We now consider the capacity of the i.i.d. Rayleigh fading MIMO channel with an arbitrary number of transmit and receive antennas. In the following proposition, we establish that the optimal Tx covariance $\bR^*$ is diagonal.

\begin{prop}
\label{prop.MIMO.general}
For the optimization problem in \eqref{eq.fading-capacity} and considering the i.i.d. Rayleigh fading MIMO channel, the optimal Tx covariance $\bR^*$ is diagonal, and Algorithm \ref{algorithm.OPA} gives the optimal solution for its diagonal entries.

\begin{proof}
Please refer to Appendix E.
\end{proof}

\end{prop}  

\section{The Simulation Result}
In this section, we present examples for the capacity for a fixed channel as well as ergodic capacity for a fading channel. Consider the diagonal channel Gram matrix $\bD(g_i)$:
\bal
\label{eq.Ex1Channel}
\{g_i\} = \{1,10,3,0.3,0.4,0.6,0.9,1,1\} 
\eal
where we have three antenna groups with corresponding per-group powers as follows:
\bal
\label{eq.Ex1PGPC}
&I(1) = \{1,2,3\}, P_1 = 2 \notag \\
&I(2) = \{4,5,6,7\}, P_2 = 12 \notag \\
&I(3) = \{8,9\}, P_3 = 4
\eal

\begin{figure}[t]
\centerline{\includegraphics[width=5in]{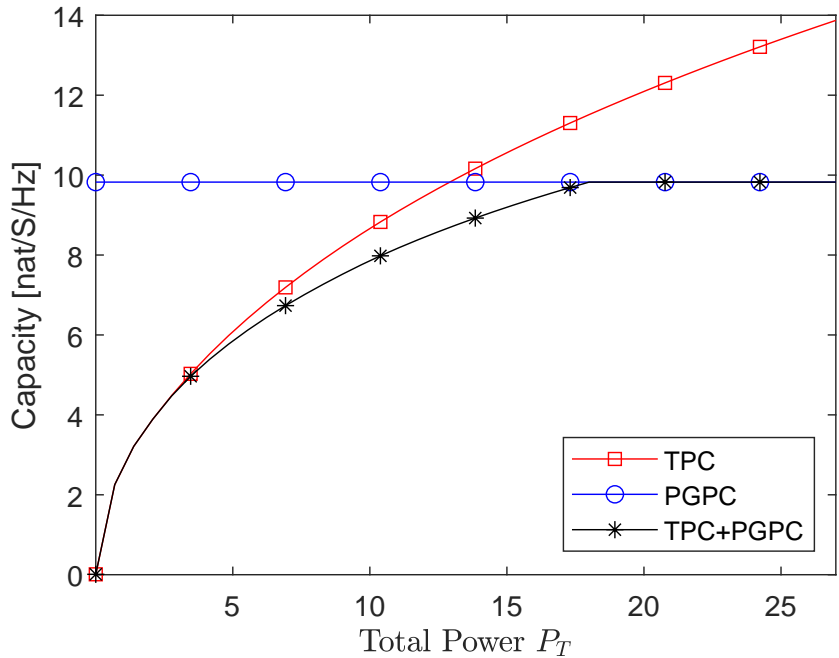}}
\caption{The capacity under the TPC, PGPC, and combined TPC and PGPC constraints for a fixed orthogonal channel in \eqref{eq.Ex1Channel}.}
\label{fig.ex1Fixed}
\end{figure}
Fig. \ref{fig.ex1Fixed}  shows the capacity under the joint power constraints where we used the solution of Theorem \ref{thm.R*-fixed}, Propositions \ref{prop.mu.lambda_j} and \ref{prop.bounds-dual}, and the bisection algorithm. It shows that the capacity under the joint TPC and PGPC is upper bounded by that under TPC alone and PGPC alone:
\bal
C^* \le \min\{C_{TPC},C_{PGPC}\}
\eal
This upper bound is tight at both low and high SNR values where $C = C_{TPC}$ and $C = C_{PGPC}$ respectively. This shows that TPC is the dominant constraint at low SNRs and PGPC is the dominant constraint at high SNRs. It can also be seen that $C_{TPC}$ is increasing in $P_T$ without bound while this is not the case for the capacity under the joint TPC and PGPC.  

Next, we consider the fading orthogonal channel for which $(H)_{ij}=0$ for $i \ne j$ and $H_{ii}$ are i.i.d. Gaussian random variables with zero mean and unit variance. Furthermore,
\bal
\label{eq.Ex2PGPC}
&I(1) = \{1,2,3\}, P_1 = 15 \notag \\
&I(2) = \{4,5,6,7\}, P_2 = 3\notag \\
&I(3) = \{8\}, P_3 = 6
\eal
\begin{figure}[t]
\centerline{\includegraphics[width=5in]{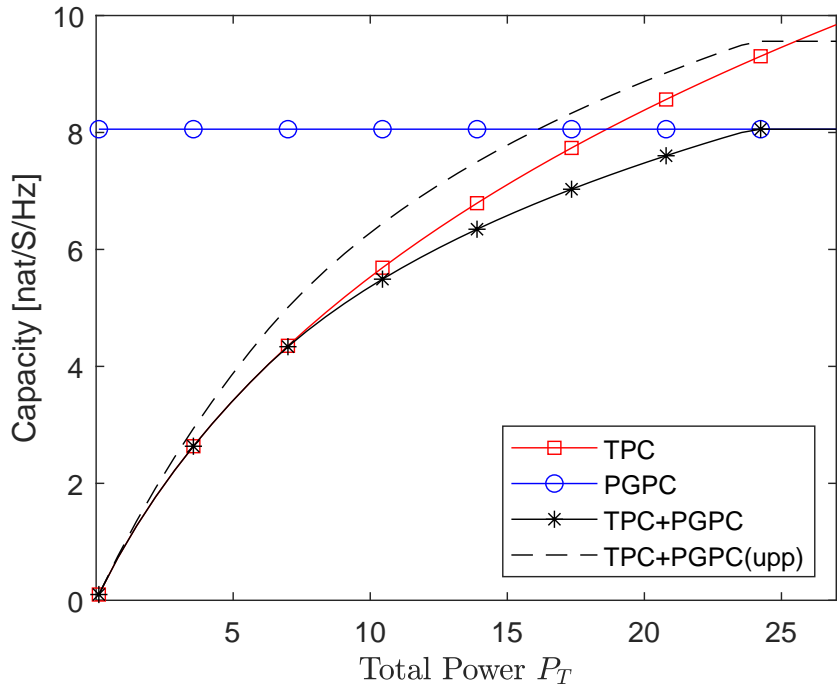}}
\caption{The ergodic capacity under the TPC, PGPC, and joint TPC and PGPC for the fading orthogonal channel and the power constraint set in \eqref{eq.Ex2PGPC}.}
\label{fig.ex2Fading}
\end{figure}
Fig. \ref{fig.ex2Fading} shows the ergodic capacity where OPA is obtained by applying the Algorithm \ref{algorithm.OPA}. In order to compute the ergodic capacity knowing the OPA, observe that:
\bal
\label{eq.Ei}
\e_{g}\big\{\ln(1+gp)\big\} &= \int_{0}^{+\infty} \ln(1+gp)e^{-g}dg \notag \\
& = - e^{(1/p)}\operatorname{Ei}(-1/p),
\eal
where $\operatorname{Ei}(\cdot)$ is the exponential integral ($\operatorname{Ei}(x) =\int_{-\infty}^{x} \frac{e^t}{t}dt $ for $x < 0 $). The first equality follows from $g = |h|^2$ and since $h$ is a Gaussian random variable with zero mean and unit variance, then $g$ has a exponential distribution with unit mean \cite{Tse-05}. To obtain \eqref{eq.Ei}, refer to page 568 of \cite{Gradshtei-00}. Again, we observe that the capacity is smaller than $C_{TPC}$ and $C_{PGPC}$. Also, the upper bound of $C$ is shown where we used the following property:
\bal
\label{eq.uppC}
\e_{g_i}\bigg\{\sum_{i=1}^m\ln(1+g_ip_i)\bigg\}  \le \sum_{i=1}^m\ln\big(1+\e_{g_i}\{g_i\}p_i\big). 
\eal
This follows from the Jensen's inequality applied to the concave function $\sum_{i=1}^m \ln(1+g_ip_i)$ \cite{Boyd-04}. The results of both Figs. \ref{fig.ex1Fixed} and \ref{fig.ex2Fading} for OPA were verified by using the Monte-Carlo simulations.

\section{Conclusion}
We studied and derived the optimal transmit covariance matrix and the capacity of the MIMO channels under the total and per-antenna group power constraints for the following cases: a) fixed and i.i.d. fading MIMO orthogonal channels, b) i.i.d. Rayleigh fading MISO channels, c) i.i.d. Rayleigh fading MIMO channels with two transmit antennas, and d) i.i.d. Rayleigh fading MIMO channel. For the fixed and i.i.d. fading MIMO channels, it is shown that the optimal transmit covariance matrix is diagonal. Using the KKT conditions, the optimal power allocations (OPA) for the transmit antennas, i.e. the diagonal entries of the optimal transmit covariance matrix, is obtained in closed-from for the fixed MIMO orthogonal channel. The result is similar to the water-filling procedure with some notable differences to point out the impact of the per-group power constraints. For the fading MIMO orthogonal channel with the i.i.d. channel gain matrix, we propose an algorithm to obtain the OPA for the transmit antennas. The proof of the algorithm's convergence to the optimal solution is based on a majorization inequality and a Schur-concavity property of the considered objective function. Finally, we demonstrated that the proposed algorithm provides the OPA for the transmit antennas for the i.i.d. Rayleigh MIMO channel. Our techniques and methodology may be extended to study similar problems.

\begin{appendices}

\section{Proof of Theorem \ref{thm.R*-fixed}}
First, we show that the optimal transmit covariance matrix $\bR^*$ is diagonal for the MIMO orthogonal channel in \eqref{eq.fixed-capacity} using the similar argument as in \cite{Telatar-99} for the case of the TPC only. Observe that:
\bal
\ln|\bI+\bW\bR| &\le \sum_{i=1}^m\ln(1+(\bW\bR)_{ii}) \notag \\
& = \sum_{i=1}^m\ln(1+g_ip_i) \notag \\
& = \ln|\bI+\bD(g_i)\bD(p_i)|,
\eal
where the inequality is due to Hadamard's inequality \cite{Zhang-11} and the first equality is due to $\bW = \bD(g_i)$ and $(\bR)_{ii} = p_i$. Also, note that $\bR \in S_R$ implies $\bD(p_i) \in S_R$ since both have the same diagonal entries. The upper bound is achieved by $\bR = \bD(p_i)$, and hence the optimal Tx covariance matrix $\bR^*$ is diagonal. 

Considering that both $\bR^*$ and $\bW$ are diagonal matrices, the optimization problem in \eqref{eq.fixed-capacity} can be simplified as follows: 
\bal
\label{eq.C*-simplified}
C^* = \max_{p_i \in S_P}\sum_{i=1}^m\ln(1+g_ip_i)
\eal
where the feasible set $S_P$ is given in \eqref{eq.S_P}. Note that we assume that $g_i >0$, since $g_i = 0$ implies that the $i$-th column of $\bH$ is a zero vector, and hence, the $i$-th Tx antenna can be removed from consideration since does not affect the rate. The problem in \eqref{eq.C*-simplified} is convex since the objective function is a concave function and all inequality constraints in $S_P$ form a convex set \cite{Boyd-04}. Also, Slater's condition holds, and hence KKT conditions are sufficient and necessary for optimality \cite{Boyd-04}. The Lagrangian for the optimization problem in \eqref{eq.C*-simplified} can be expressed as follows:
\bal
\label{eq.Lagrange}
L = - \sum_{i=1}^m\ln(1+g_ip_i) + \mu \Big(\sum_{i=1}^mp_i - P_T\Big) + \sum_{j=1}^s\bigg(\lambda_j\Big(\sum_{i \in I(j)}p_i-P_j\Big)\bigg) - \sum_{i=1}^m\eta_i p_i
\eal
where $\mu,\lambda_j,\eta_i \ge 0$ are the Lagrange multipliers responsible for TPC, PGPC, and $p_i \ge 0$ constraints, respectively. Using the  Lagrangian and the feasible set, the KKT conditions for the optimality can be expressed as follows:
\bal
\label{eq.KKT1}
&\frac{\partial L} {\partial p_i}= -\frac{g_i}{1+g_ip_i}+\mu + \lambda_j -\eta_i=0,\ i \in I(j)\\
\label{eq.KKT2}
&\mu\Big(\sum_{i=1}^m p_i-P_T\Big)=0, \ \lambda_j\Big(\sum_{i \in I(j)}p_i-P_j\Big) =0, \ \eta_i p_i = 0 \\
\label{eq.KKT3}
& \mu,\lambda_j,\eta_i \ge 0 \\
\label{eq.KKT4}
& p_i \ge 0,\ \sum_{i=1}^mp_i \le P_T,\ \sum_{i \in I(j)}p_i \le P_j
\eal
where \eqref{eq.KKT1}-\eqref{eq.KKT4} correspond to: (i) stationary condition, (ii) complementary slackness condition, (iii) dual feasibility constraints, and (iv) primal feasibility constraints, respectively. From \eqref{eq.KKT1}, we observe that:
\bal
\label{eq.p_i}
p_i = (\mu+\lambda_j-\eta_i)^{-1}-g_i^{-1}
\eal
Next, we simplify the equation in \eqref{eq.p_i} by considering two cases of $p_i >0$ and $p_i = 0$. First, we consider $p_i >0$ which implies $\eta_i = 0$ by using \eqref{eq.KKT2}. For this case, \eqref{eq.p_i} implies:
\bal
\label{eq.pi>0}
0<p_i = (\mu+\lambda_j)^{-1}-g_i^{-1}
\eal
Therefore, $p_i$ is as in \eqref{eq.pi>0} if:
\bal
\label{eq.pi>0-constraint}
(\mu+\lambda_j)^{-1}>g_i^{-1}
\eal
For the case of $p_i = 0$, we observe that from \eqref{eq.KKT3}, \eqref{eq.p_i}:
\bal
\label{eq.pi=0}
0 \le \eta_i = (\mu+\lambda_j) - g_i 
\eal
Thus, $p_i = 0$ if:
\bal
(\mu+\lambda_j)^{-1}\le g_i^{-1}
\eal
One can combine these two cases and shows the following expression for $p_i$:
\bal
\label{eq.combinded-p_i}
p_i= \begin{cases} 0 & \quad (\mu+\lambda_j)^{-1}\le g_i^{-1} \\
(\mu+\lambda_j)^{-1}-g_i^{-1} & \quad (\mu+\lambda_j)^{-1}>g_i^{-1} 
\end{cases}
\eal
This solution for the optimal power allocation can be expressed as $p^*_i = \big((\mu+\lambda_j)^{-1}-g_i^{-1}\big)_+$.

\section{Proof of Proposition \ref{prop.mu.lambda_j}}
\subsection{Proof for the calculation of dual variable $\mu$}If $\sum_{j=1}^s P_j \le P_T$, then $\sum_{i=1}^m p_i \le P_T$ and TPC is inactive ($\mu = 0$ follows from  \eqref{eq.KKT2}\footnote{In the case of $\sum_{j=1}^s P_j \le P_T$, the total power constraint becomes redundant as removing it, i.e. $\mu=0$, does not change the feasible set of the problem.}). Otherwise, $\mu > 0$ and the TPC is active $\sum_{i=1}^m p_i = P_T$. In this case, we consider two possible cases of $\lambda_j > 0 $ and $\lambda_j = 0$. $\lambda_j >0$ implies:
\bal
\label{eq.lambda_j >0-1}
P_j  &= \sum_{i \in I(j)}p_i \\
\label{eq.lambda_j >0-2}
&=\sum_{i \in I(j)}\big((\mu+\lambda_j)^{-1}-g_i^{-1}\big)_+ \\
\label{eq.lambda_j >0-3}
& \le \sum_{i \in I(j)}(\mu^{-1}-g_i^{-1})_+  
\eal
where \eqref{eq.lambda_j >0-1} follows from the complementary slackness condition in \eqref{eq.KKT2}, equality in \eqref{eq.lambda_j >0-2} is due to Theorem \ref{thm.R*-fixed}, the inequality in \eqref{eq.lambda_j >0-3} follows from $\lambda_j > 0$. 

For the case of $\lambda_j =0$, Theorem \ref{thm.R*-fixed} implies that:
\bal
\label{eq.lambda_j =0-1}
\sum_{i \in I(j)}p_i  = \sum_{i \in I(j)}(\mu^{-1}-g_i^{-1})_+.
\eal
Observe that:
\bal
\label{eq.lambda_j =0-2}
P_j &\ge \sum_{i \in I(j)}p_i  \\
\label{eq.lambda_j =0-3}
&=\sum_{i \in I(j)}(\mu^{-1}-g_i^{-1})_+ & 
\eal
where \eqref{eq.lambda_j =0-2} follows from the primal feasibility constraints in \eqref{eq.KKT4} and \eqref{eq.lambda_j =0-3} is due to \eqref{eq.lambda_j =0-1}.

Combining both cases, one obtains:
\bal
\label{eq.combinded-P^_j}
\sum_{i \in I(j)}p_i = \begin{cases} P_j,  &\sum_{i \in I(j)}(\mu^{-1}-g_i^{-1})_+ \ge  P_j \\
\sum_{i \in I(j)}(\mu^{-1}-g_i^{-1})_+,  &\sum_{i \in I(j)}(\mu^{-1}-g_i^{-1})_+ \le  P_j
\end{cases}
\eal
Hence, 
\bal
\label{eq.sum_I(j)p_i}
\sum_{i \in I(j)}p_i = \min\bigg\{P_j, \sum_{i \in I(j)}(\mu^{-1}-g_i^{-1})_+\bigg\}
\eal
Combining \eqref{eq.sum_I(j)p_i} with the fact that $\sum_{j=1}^s\sum_{i \in I(j)}p_i =\sum_{i=1}^m p_i = P_T$ yields \eqref{eq.mu.find}.

\subsection{Proof for the calculation of dual variable $\lambda_j$}Once $\mu$ is obtained, $\sum_{i \in I(j)}p_i $ can be found using \eqref{eq.sum_I(j)p_i}. Also,
\bal
\label{eq.sum_I(j)p_i.lambda_j}
\sum_{i \in I(j)}p_i =\sum_{i \in I(j)}\big((\mu+\lambda_j)^{-1}-g_i^{-1}\big)_+
\eal
where this equality is due to Theorem \ref{thm.R*-fixed}. Combining \eqref{eq.sum_I(j)p_i} with \eqref{eq.sum_I(j)p_i.lambda_j} yields \eqref{eq.lambda_j.find}.

\section{Proof of Proposition \ref{prop.algorithm}}

Below, we state the definition of majorization inequality and its application to Schur-concave functions.
\begin{definition}
\label{remark.maj}
Let $\bx = \big[x[1],x[2],...,x[m]\big]$ and $\by = \big[y[1],y[2],...,y[m]\big]$; both are in decreasing order. Then, $\bx$ is majorized by $\by$ if:
\bal
\sum_{i=1}^{k} x[i]&\le \sum_{i=1}^{k} y[i],\ k=1,2,\ldots,m-1, \notag \\
\sum_{i=1}^{m} x[i ]&= \sum_{i=1}^{m} y[i].
\eal
where the majorization inequality is noted as $\bx \prec \by$. Then, $f(\cdot)$ is a Schur-concave function if $\bx \prec \by$ implies $f(\bx) \ge f(\by)$ \cite{Zhang-11} \cite{Pecaric-92}.
\end{definition}

To prove the proposition, note that this algorithm converges since there is the finite number of groups, and as well the values in $S_P$ are finite. Let us consider that there are $t$ active groups when the algorithm converges. We arrange them in increasing order of $\frac{P_u}{|I(u)|}$:
\bal
\label{eq.inc.hat(P_j)/I(j)}
\frac{P_{j_1}}{|I(j_1)|} \le \frac{P_{j_2}}{|I(j_2)|} \le ... \le \frac{P_{j_t}}{|I(j_t)|}.
\eal 
Note that in the algorithm, $\frac{P}{L}$ increases in each iteration. This follows from the fact that $\frac{P}{L} \ge \frac{P_u}{|I(u)|}$ implies $\frac{P-P_u}{L-|I(u)|}  \ge \frac{P}{L}$ ($P\ge  P_u >0, L> |I(u)| > 0 $). Hence, after the last step:
\bal
\label{eq.P/L>hat(P_jt)/I(jt)}
\frac{P_{j_t}}{|I(j_t)|} \le \frac{P}{L} = \frac{P_T-\sum_{k=1}^tP_{j_k}}{m-\sum_{k=1}^t|I(j_k)|}.
\eal
Eq. \eqref{eq.inc.hat(P_j)/I(j)} and \eqref{eq.P/L>hat(P_jt)/I(jt)} imply that:
\bal
\frac{P_{j_1}}{|I(j_1)|} \le ... \le \frac{P_{j_t}}{|I(j_t)|} \le \frac{P_T-\sum_{k=1}^tP_{j_k}}{m-\sum_{k=1}^t|I(j_k)|} 
\eal
In this case, the power allocation $\bpp^* = \{p^*_i\}$ for antennas is as follows:
\bal
\label{eq.bp^*}
\bpp^* = \Bigg[{\underbrace{\frac{P_{j_1}}{|I(j_1)|}, .., \frac{P_{j_1}}{|I(j_1)|}}_{\text{antennas in the 1st active group}}} ,...,{\underbrace{\frac{P_{j_t}}{|I(j_t)|},...,\frac{P_{j_t}}{|I(j_t)|}}_{\text{antennas in the $t$-th active group}}}, {\underbrace{\frac{P_T-\sum_{k=1}^tP_{j_k}}{m-\sum_{k=1}^t|I(j_k)|} ,..., \frac{P_T-\sum_{k=1}^tP_{j_k}}{m-\sum_{k=1}^t|I(j_k)|}}_{\text{antennas in the inactive groups}}} \Bigg]
\eal
Note that $\bpp^*$ is feasible under the joint TPC and PGPC constraints, since it is found from Algorithm \ref{algorithm.OPA}, and both the TPC and PGPC constraints are considered in the algorithm. If the power allocation $\bpp^*$ is not optimal, the optimal solution has the following general form:\\
(i) If $1\le k \le t$, the $k$-th group power is $P_{j_k} - \epsilon_{j_k}$ where $\epsilon_{j_k} \ge 0$;\\
(ii) if $k>t$, the remaining total power for the rest of the antennas in the rest of groups is $P_T-\sum_{k=1}^tP_{j_k} + \sum_{k=1}^t \epsilon_{j_k}$.

We need to obtain the maximum achievable rate considering this setting of power allocation. In this case, the maximum achievable rate is achieved when all antennas in a particular group get the same amount of power (since the objective function is symmetric and concave \cite{Tse-05}). Hence, the maximum achievable rate under this setting is achieved for the following power allocation:
\bal
\label{eq.contradict.active}
&i \in I(j_k), 1\le k \le t: p^{**}_i=\frac{P_{j_k} - \epsilon_{j_k}}{I(j_k)} \\
\label{eq.contradict.inactive}
&i \in I(j_k), k > t: p^{**}_i = \frac{P_T-\sum_{k=1}^tP_{j_k} + \sum_{k=1}^t \epsilon_{j_k}}{m-\sum_{k=1}^t|I(j_k)|}
\eal
Next, let's arrange $\bpp^{**} = \{p^{**}_i\}$ in increasing order:
\bal
\bpp^{**} = \Bigg[
\frac{P_{i_1}-\epsilon_{i_1}}{|I(i_1)|}, .., \frac{P_{i_1}-\epsilon_{i_1}}{|I(i_1)|},...,\frac{P_{i_t}-\epsilon_{i_t}}{|I(i_t)|},...,\frac{P_{i_t}-\epsilon_{i_t}}{|I(i_t)|}, \frac{P_T-\sum_{k=1}^tP_{j_k} + \sum_{k=1}^t \epsilon_{j_k}}{m-\sum_{k=1}^t|I(j_k)|},..., \notag\\
\label{eq.bp^**}
\frac{P_T-\sum_{k=1}^tP_{j_k} + \sum_{k=1}^t \epsilon_{j_k}}{m-\sum_{k=1}^t|I(j_k)|}
\Bigg]
\eal
Note that we change the notation from $P_{j_k}$ to $P_{i_k}$\footnote{In the proof of this proposition, we have two different ordering sequences: $P_{j_k}$ and $P_{i_k}$. For the former sequence, $\frac{P_{j_k}}{|I(j_k)|}$ is in increasing order and for the latter sequence, $\frac{P_{i_k}-\epsilon_{i_k}}{|I(i_k)|}$ is in increasing order.}. 
In order to find the optimal solution for $p_i$, we need to compare $C(\bpp^*)$ with $C(\bpp^{**})$. To do this comparison, we use the following arguments:

1- $C(\bpp)$ is a Schur-concave function (see Lemma \ref{lemma.Schur-concave}).

2- $\bpp^*$ is majorized by $\bpp^{**}$ (see Lemma \ref{lemma.majorization}). \\
Hence:
\bal
\bpp^{**}\succ \bpp^* \Rightarrow C(\bpp^*) \ge C(\bpp^{**})
\eal 
where it shows that $\bpp^*$ gives the OPA for $p_i$.

\begin{lemma}
\label{lemma.Schur-concave}
The function $C(\bpp=\{p_i\})$ in \eqref{eq.C(p_i)} is Schur-concave.

\begin{proof}
The function $C(\bpp=\{p_i\}) = \e_{g_i}\bigg\{\sum_{i=1}^m\ln(1+g_ip_i)\bigg\} $ is permutation-invariant (symmetric) function \cite{Tse-05}. Also, it is concave function, which follows from the fact that $\bigg\{\sum_{i=1}^m\ln(1+g_ip_i)\bigg\}$ is a concave function in $p_i$ and expectation $ \e_{g_i}$ is a linear function (it preserves the inequality). Permutation-invariant property and concavity imply that this is a Schur-concave function (see Theorems 12.27 and 12.25 in \cite{Pecaric-92}) 
\end{proof}

\end{lemma}

\begin{lemma}
\label{lemma.majorization}
The vector $\bpp^*$ is majorized by $\bpp^{**}$.

\begin{proof}
 Let's rearrange $\bpp^*$ and $\bpp^{**}$ in decreasing order, and express them in the more general form:
\bal
\label{eq.p*Increasing}
&\bpp^{*\downarrow} = \bx_1=[{\underbrace{x,...,x}_{\text{$q$ elements of $x$}}},x_{q+1},...,x_m] \\
\label{eq.p**Increasing}
&\bpp^{**\downarrow} = \bx_2 = \bigg[{\underbrace{x+\frac{1}{q}\sum_{i=q+1}^m\epsilon_i,...,x+\frac{1}{q}\sum_{i=q+1}^m\epsilon_i}_{\text{$q$ elements of $x+\frac{1}{q}\sum_{i=q+1}^m\epsilon_i$}}}, (x-\epsilon)_{q+1},\ldots,(x-\epsilon)_{m} \bigg]
\eal
where $\epsilon_i \ge 0$ for all $i$ and $(x-\epsilon)_{t} \ge (x-\epsilon)_{t+1}$, i.e. we arrange $(x_i-\epsilon_i)$ in decreasing order for $\bx_2$ (e.g. $(x-\epsilon)_{q+1}$ and $(x-\epsilon)_{q+2}$ are the largest and the second largest entries of $\{(x_{q+1}-\epsilon_{q+1}),...,(x_{m}-\epsilon_{m})\}$). Eq. \eqref{eq.p*Increasing} is a more general representation of \eqref{eq.bp^*} where there are $q$ antennas in inactive groups and the total number of antennas is $m$. To obtain \eqref{eq.p**Increasing}, the antennas' powers in active groups, i.e. $x_{q+1},...,x_m$, are deducted by $\epsilon_i$, and $\frac{1}{q}\sum_{i=q+1}^m\epsilon_i$ is added to all antennas' powers in inactive groups (see the relation of \eqref{eq.bp^**} with \eqref{eq.bp^*}). Here, we denote the $i$-th element of the vector $\bx_1$ with $\bx_1[i]$. Observe that:
\bal
\label{eq.maj.=}
\sum_{i=1}^m\bx_2[i] = \sum_{i=1}^m\bx_1[i]
\eal
Also,
\bal
\label{eq.maj.n<=q}
\text{for}~~n\le q: \sum_{i=1}^n\bx_2[i] &= nx+\frac{n}{q} \sum_{i=q+1}^m\epsilon_i \overset{(a)}{\ge} nx = \sum_{i=1}^n\bx_1[i] \\
\text{for}~~n> q: \sum_{i=1}^n\bx_2[i] &= qx+\sum_{i=q+1}^m\epsilon_i + \sum_{i=q+1}^n(x-\epsilon)_{i} \notag \\
& \overset{(b)}{\ge}  qx+\sum_{i=q+1}^m\epsilon_i + \sum_{i=q+1}^n(x_i-\epsilon_i) \notag \\
& =  qx+\sum_{i=q+1}^n x_i+\sum_{i=n+1}^m\epsilon_i \notag \\
\label{eq.maj.n>q}
&\overset{(c)}{\ge} qx+\sum_{i=q+1}^n x_i  = \sum_{i=1}^n\bx_1[i]
\eal
where (a) and (c) are due to $\epsilon_i \ge 0$ for all $i$, (b) follows from the fact that $\sum_{i=q+1}^n(x-\epsilon)_{i}$ is the summation of the first $n-q$ largest values of $\{(x_{q+1}-\epsilon_{q+1}),...,(x_{m}-\epsilon_{m})\}$. Then, \eqref{eq.maj.n<=q} and \eqref{eq.maj.n>q} imply that $\sum_{k=1}^n\bx_2[k] \ge \sum_{k=1}^n\bx_1[k]$. Combining this with \eqref{eq.maj.=}, we observe that $\bx_2 \succ \bx_1$.

\end{proof}

\end{lemma}

\section{Proof of Proposition \ref{prop.2.n.MIMO}}

Let  $\bR,\bW \ge 0$ be as follows:
\bal
\label{eq.2.R.Z}
\bR = \left[
\begin{array}{cc}
   r_1& r_3\\
   r^*_3 & r_2 \\
\end{array}
\right],\bW = \left[
\begin{array}{cc}
   w_1& w_3\\
   w^*_3 & w_2 \\
\end{array}
\right]. 
\eal
To prove this proposition, we need the following lemmas. 

\begin{lemma}
\label{lemma.w3}
Let the entries of $\bH_{n \times 2}$ are i.i.d. complex Gaussian with zero mean and $\bW_{2 \times 2} = \bH^+\bH$. Then, $f_{w_3}(w_3) = f_{w_3}(-w_3)$, where $f_{w_3}(w_3)$ is the probability density function (pdf) of $w_3$.

\begin{proof}
Observe that $w_3$ is as follows:
\bal
w_3 = \sum_{i=1}^n(\bH)^*_{1i}(\bH)_{2i}.
\eal
Note that $(\bH)^*_{1i}$ and $(\bH)_{2i}$ are independent and have the identical complex Gaussian distribution with zero mean. It implies that for the random variable $x_i=(\bH)^*_{1i}(\bH)_{2i}$ (i.e. the product of two independent complex Gaussian random variables with zero mean), we observe $f_{x_i}(x_i) = f_{x_i}(-x_i)$, where $f_{x_i}(x_i)$ is the pdf of $x_i$ \cite{ODonoughue-12}\cite{Simon-02}\cite{Betlehem-12}. Also,
\bal
\label{eq.conv}
f_{w_3}(-w_3) = f_{x_1}(-w_3)  \circledast   f_{x_2}(-w_3) \circledast ... \circledast   f_{x_n}(-w_3)
\eal
where $\circledast$ is the two dimensional convolution operator including both real and imaginary part, since all random variables are complex. The equality follows from the fact that the pdf of sum of independent random variables equals the convolution of the pdf of those random variables \cite{Papoulis-01}, as it is shown above. Hence, \eqref{eq.conv} implies that $f_{w_3}(w_3) = f_{w_3}(-w_3)$ since $f_{x_i}(w_3) = f_{x_i}(-w_3)$.

\end{proof}

\end{lemma}

\begin{lemma}
\label{lemma.diagonalR}
Consider $2 \times n$ MIMO channel. The optimization problem in \eqref{eq.fading-capacity} is upper bounded as follows:
\bal
\label{eq.upper.C*}
C^* \le  \max_{\bR \in S_R}\e_{w_i} \Big\{\ln \big(r_1r_2(w_1w_2-|w_3|^2)+w_1r_1+w_2r_2+1\big) \Big\}
\eal
and the upper-bound is achieved when $r_3 = 0$.
\begin{proof}
For $\bH_{2 \times n}$, the objective function in \eqref{eq.fading-capacity} is upper bounded as follows:
\bal
\e_{H}\big\{\ln &|\bI+\bH\bR\bH^+|\big\} \overset{(a)}{=} \e_{W}\big\{\ln |\bI+\bW\bR|\big\} \notag \\
&=\e_{w_i} \bigg\{ \ln\Big((r_1r_2-|r_3|^2)(w_1w_2-|w_3|^2)+w_1r_1+w_2r_2+1 + 2 \Re(w_3r^*_3)\Big)\bigg\} \notag \\
 &\overset{(b)}{\le} \e_{w_i} \bigg\{ \ln\Big(r_1r_2(w_1w_2-|w_3|^2)+w_1r_1+w_2r_2+1 + 2 \Re(w_3r^*_3)\Big)\bigg\} \notag \\
&\overset{(c)}{=} \frac{1}{2} \e_{w_i} \bigg\{ \ln\Big(r_1r_2(w_1w_2-|w_3|^2)+w_1r_1+w_2r_2+1 + 2 \Re(w_3r^*_3)\Big) \notag \\ 
&\quad \quad \quad \ + \ln \Big(r_1r_2(w_1w_2-|w_3|^2)+w_1r_1+w_2r_2+1 - 2\Re(w_3r^*_3)\Big)\bigg\} \notag \\
&= \frac{1}{2} \e_{w_i} \bigg\{ \ln\Big(\big(r_1r_2(w_1w_2-|w_3|^2)+w_1r_1+w_2r_2+1\big)^2 - 4 \big(\Re(w_3r^*_3)\big)^2\Big) \bigg\} \notag \\
&\overset{(d)}{\le} \frac{1}{2} \e_{w_i} \bigg\{ \ln\big(r_1r_2(w_1w_2-|w_3|^2)+w_1r_1+w_2r_2+1\big)^2 \bigg\} \notag \\
& = \e_{w_i} \bigg\{ \ln\big(r_1r_2(w_1w_2-|w_3|^2)+w_1r_1+w_2r_2+1\big) \bigg\}
\eal 
where (a) is due to $|\bI+\bA\bB| = |\bI+\bB\bA|$, (b) is due to $r_1r_2 \ge |r_3|^2, \ w_1w_2 \ge |w_3|^2$, (c) follows from Lemma \ref{lemma.w3} since changing the sign of $w_3$ does not affect the expectation, (d) is due to $\big(\Re(w_3r^*_3)\big)^2 \ge 0$. Taking the $\max$ operator of both sides leads \eqref{eq.upper.C*}. The upper bound is achieved when $r_3=0$.
\end{proof}

\end{lemma}  

Lemma \ref{lemma.w3} shows that the channel capacity is achieved for $r_3=0$, i.e. $\bR^* =\bD(r_1,r_2) \overset{\bigtriangleup}{=} \bD(p_1,p_2)$ is optimal. The objective function in \eqref{eq.fading-capacity} can be simplified to $\e_{H}\big\{\ln |\bI+\bH \bD(r_i)\bH^+|\big\}$. For the i.i.d. Rayleigh fading, this function is symmetric in $r_i$ and also jointly concave \cite{Tse-05}, and hence is Schur-concave \cite{Pecaric-92}. Hence, using the same argument for the proof of the Proposition \ref{prop.algorithm}, we can show that Algorithm \ref{algorithm.OPA} gives the OPA for $p_i$. In this case, either there are two antenna groups, i.e. $I(1) = {1}$ and $I(2) = {2}$, or there is only one group, i.e. $I(1) = {1,2}$. For the latter case, the problem becomes finding the capacity of the MIMO channel under the total power constraint only. 

\section{Proof of Proposition \ref{prop.MIMO.general}}
Consider the i.i.d Rayleigh fading MIMO channel under joint TPC and PGPC constraints. We denote the $i$-th eigenvalue and the $i$-th diagonal entries of $\bR$ by $\lambda_i$ and $d_i$ respectively. We observe:
\bal
\label{fading.MIMO.general.proof1}
\e_{\bH}\big\{\ln|\bI+\bH\bR\bH^+|\big\} &= \e_{\bH}\big\{\ln|\bI+\bH\bU\bD(\lambda_i)\bU^+\bH^+|\big\} \\
\label{fading.MIMO.general.proof2}
& = \e_{\bH}\big\{\ln|\bI+\bH\bD(\lambda_i)\bH^+|\big\} \\
\label{fading.MIMO.general.proof3}
& \le \e_{\bH}\big\{\ln|\bI+\bH\bD(d_i)\bH^+|\big\}
\eal
In \eqref{fading.MIMO.general.proof1}, we use the eigenvalue decomposition of $\bR$, i.e. $\bR = \bU\bD(\lambda_i)\bU^+$, where the diagonal entries of the diagonal matrix $\bD(\lambda_i)$ are the eigenvalues of $\bR$, and $\bU$ is the unitary matrix contains the eigenvectors of $\bR$ in its columns; \eqref{fading.MIMO.general.proof2} is due to the fact that $\bH$ and $\bH\bU$ have the identical distribution for the unitary matrix $\bU$ and for the i.i.d. Rayleigh fading MIMO channel $\bH$ \cite{Tse-05}; to show \eqref{fading.MIMO.general.proof3}, note that: i) $f(\lambda_i) = \e_{\bH}\big\{\ln|\bI+\bH\bD(\lambda_i)\bH^+|\big\}$ is the Schur-concave function since it is the concave function and is symmetric in $\lambda_i$ \cite{Tse-05} (see also Lemma \ref{lemma.Schur-concave}), ii) the eigenvalues of a Hermitian matrix majorizes its diagonal entries \cite{Zhang-11}, which implies that $[\lambda_1, \lambda_2,\ldots, \lambda_m] \succ [d_1, d_2,\ldots, d_m]$. Therefore, $f(\lambda_i) \le f(d_i)$ which follows from these two facts. Since $\bR$ and $\bD(d_i)$ have the same diagonal entries, it follows that $\bD(d_i) \in S_R$. Hence, $\bR^*$ has to be a diagonal matrix\footnote{Note that $\bR$ in \eqref{fading.MIMO.general.proof1} and $\bD(d_i)$ in \eqref{fading.MIMO.general.proof3} have the same diagonal entries. This can be achieved by a proper reordering of $d_i$ in \eqref{fading.MIMO.general.proof3} since $\e_{\bH}\big\{\ln|\bI+\bH\bD(d_i)\bH^+|\big\}$ is symmetric in $d_i$.}. 
Now, the channel capacity in \eqref{eq.fading-capacity} can be written as follows
\bal
C^* = \max_{p_i \in S_P}\e_{\bH}\big\{\ln|\bI+\bH\bD(p_i)\bH^+|\big\}
\eal
Here, the objective function is Schur-concave, and therefore Algorithm 1 gives the optimal solution for $p_i$ following the proof given in Proposition \ref{prop.algorithm}.

\end{appendices}

\bibliographystyle{IEEEtran} 
\bibliography{IEEEabrv,main}

\end{document}